\documentstyle[12pt]{article}
\newlength{\pubnumber} 

\catcode`\@=11
\@addtoreset{equation}{section}

\def\section{\@startsection{section}{1}{\z@}{3.5ex plus 1ex minus .2ex}
 {2.3ex plus .2ex}{\large\bf}}
\def\subsection{\@startsection{subsection}{2}{\z@}{2.3ex plus .2ex}
 {2.3ex plus .2ex}{\bf}}

\def\LG{Landau--Ginzburg}
\begin{document}

\begin{titlepage}
\samepage{
\setcounter{page}{1}
\rightline{ACT--4/98}
\rightline{CPT--TAMU--12/98}
\rightline{NSF--ITP--98--29}
\rightline{UFIFT-HEP-98--9}
\rightline{\tt hep-th/9803262}
\rightline{March 1998}
\vfill
\begin{center}
 {\Large \bf Toward the $M$($F$)--Theory Embedding of \\
Realistic Free-Fermion Models \\}
\vfill
\vspace{.25in}
 {\large P. Berglund$^{1}$, J. Ellis$^{2}$,
   A.E. Faraggi$^{3}$, D.V. Nanopoulos$^{4}$,
   $\,$and$\,$ Z. Qiu$^3$\\}
\vspace{.25in}
{\it $^{1}$ Institute for Theoretical Physics, University of California,
     Santa Barbara, CA~93106, USA\\}
\vspace{.05in}
 {\it  $^{2}$ Theory Division, CERN, CH-1211 Geneva, Switzerland \\}
\vspace{.05in}
 {\it  $^{3}$ Institute for Fundamental Theory, Department of Physics,\\
              University of Florida, Gainesville, FL  32611, USA\\}
\vspace{.05in}
 {\it  $^{4}$ Center for
Theoretical Physics, Dept. of Physics,
Texas A \& M University, College Station, TX~77843-4242, USA,  \\
Astroparticle Physics Group, Houston
Advanced Research Center (HARC), The Mitchell Campus,
Woodlands, TX~77381, USA, and \\
Academy of Athens, Chair of Theoretical Physics,
Division of Natural Sciences, 28~Panepistimiou Avenue,
Athens 10679, Greece.\\}
\end{center}
\vfill
\begin{abstract}
  {\rm
We construct a \LG~model with the same data and symmetries
as a $Z_2\times Z_2$ orbifold that corresponds to a
class of realistic free-fermion models.
Within the class of interest, we show that this orbifolding
connects between different $Z_2\times Z_2$ orbifold models
and commutes with the mirror symmetry. Our work suggests
that duality symmetries previously discussed in the context of
specific $M$ and $F$ theory compactifications
may be extended to the special $Z_2\times Z_2$ 
orbifold that characterizes realistic free-fermion models.
}
\end{abstract}
\vfill
\smallskip}
\end{titlepage}

\setcounter{footnote}{0}

\def\beq{\begin{equation}}
\def\eeq{\end{equation}}
\def\beqn{\begin{eqnarray}}
\def\eeqn{\end{eqnarray}}
\def\Tr{{\rm Tr}\,}
\def\KM{{Ka\v{c}-Moody}}

\def\ie{{\it i.e.}}
\def\etc{{\it etc}}
\def\eg{{\it e.g.}}
\def\half{{\textstyle{1\over 2}}}
\def\third{{\textstyle {1\over3}}}
\def\quarter{{\textstyle {1\over4}}}
\def\m{{\tt -}}
\def\p{{\tt +}}

\def\rep#1{{\bf {#1}}}
\def\slash#1{#1\hskip-6pt/\hskip6pt}
\def\slk{\slash{k}}
\def\GeV{\,{\rm GeV}}
\def\TeV{\,{\rm TeV}}
\def\y{\,{\rm y}}
\def\SM{Standard-Model }
\def\SUSY{supersymmetry }
\def\SSM{supersymmetric standard model}
\def\vev#1{\left\langle #1\right\rangle}
\def\l{\langle}
\def\r{\rangle}

\def\Htw{{\tilde H}}
\def\chibar{{\overline{\chi}}}
\def\qbar{{\overline{q}}}
\def\ibar{{\overline{\imath}}}
\def\jbar{{\overline{\jmath}}}
\def\Hbar{{\overline{H}}}
\def\Qbar{{\overline{Q}}}
\def\abar{{\overline{a}}}
\def\alphabar{{\overline{\alpha}}}
\def\betabar{{\overline{\beta}}}
\def\tautwo{{ \tau_2 }}
\def\calF{{\cal F}}
\def\calP{{\cal P}}
\def\calN{{\cal N}}
\def\smallmatrix#1#2#3#4{{ {{#1}~{#2}\choose{#3}~{#4}} }}
\def\bone{{\bf 1}}
\def\V{{\bf V}}
\def\b{{\bf b}}
\def\N{{\bf N}}
\def\bQ{{\bf Q}}
\def\t#1#2{{ \Theta\left\lbrack \matrix{ {#1}\cr {#2}\cr }\right\rbrack }}
\def\C#1#2{{ C\left\lbrack \matrix{ {#1}\cr {#2}\cr }\right\rbrack }}
\def\tp#1#2{{ \Theta'\left\lbrack \matrix{ {#1}\cr {#2}\cr }\right\rbrack }}
\def\tpp#1#2{{ \Theta''\left\lbrack \matrix{ {#1}\cr {#2}\cr }\right\rbrack }}


\def\inbar{\,\vrule height1.5ex width.4pt depth0pt}

\def\IC{\relax\hbox{$\inbar\kern-.3em{\rm C}$}}
\def\IQ{\relax\hbox{$\inbar\kern-.3em{\rm Q}$}}
\def\IR{\relax{\rm I\kern-.18em R}}
 \font\cmss=cmss10 \font\cmsss=cmss10 at 7pt
\def\IZ{\relax\ifmmode\mathchoice
 {\hbox{\cmss Z\kern-.4em Z}}{\hbox{\cmss Z\kern-.4em Z}}
 {\lower.9pt\hbox{\cmsss Z\kern-.4em Z}}
 {\lower1.2pt\hbox{\cmsss Z\kern-.4em Z}}\else{\cmss Z\kern-.4em Z}\fi}

\def\AEF{A.E. Faraggi}
\def\KRD{K.R. Dienes}
\def\JMR{J. March-Russell}
\def\NPB#1#2#3{{\it Nucl.\ Phys.}\/ {\bf B#1} (19#2) #3}
\def\PLB#1#2#3{{\it Phys.\ Lett.}\/ {\bf B#1} (19#2) #3}
\def\PRD#1#2#3{{\it Phys.\ Rev.}\/ {\bf D#1} (19#2) #3}
\def\PRL#1#2#3{{\it Phys.\ Rev.\ Lett.}\/ {\bf #1} (19#2) #3}
\def\PRT#1#2#3{{\it Phys.\ Rep.}\/ {\bf#1} (19#2) #3}
\def\MODA#1#2#3{{\it Mod.\ Phys.\ Lett.}\/ {\bf A#1} (19#2) #3}
\def\IJMP#1#2#3{{\it Int.\ J.\ Mod.\ Phys.}\/ {\bf A#1} (19#2) #3}
\def\nuvc#1#2#3{{\it Nuovo Cimento}\/ {\bf #1A} (#2) #3}
\def\etal{{\it et al,\/}\ }

\hyphenation{su-per-sym-met-ric non-su-per-sym-met-ric}
\hyphenation{space-time-super-sym-met-ric}
\hyphenation{mod-u-lar mod-u-lar--in-var-i-ant}


\setcounter{footnote}{0}
\section{Introduction}

The rapid recent development of tools based on
duality for the analysis of superstring models has
opened up a new range of possibilities beyond the
weak-coupling models studied earlier. Moreover, it has been
suggested that matching the bottom-up estimation of
the unification scale based on low-energy data with the
top-down estimate based on string theory is easiest in a
strong-coupling limit of heterotic string \cite{witten}.
These developments prompt a reformulation of the string
model-building approaches pioneered previously.

Among phenomenological superstring models,
those~\cite{fsu5,fny,alr,eu,lykken} derived
in the free--fermion formulation~\cite{fff} have been
taken the furthest in attempts to construct a 
semi--realistic string model that can incorporate
the Standard Model of particle physics~\cite{ffmreviews}.
This may reflect the fact that the true string
vacuum, even if strongly--coupled and still elusive, may share some
properties with the models constructed
thus far. In order to appreciate the realistic nature
of the free-fermion models, it is useful to
recall their underlying structure. 
These free--fermion models may be phrased in terms of
an underlying $Z_2\times Z_2$
orbifold compactification, by going to the point
in the Narain moduli space \cite{Narain} where all the 
internal compactified dimensions
can be represented as free world--sheet fermions.
Three--generation models then arise naturally,
due to the structure of the $Z_2\times Z_2$
orbifold. Furthermore, an important
property of the free--fermion models
is the standard $SO(10)$ embedding of the weak
hypercharge, which ensures natural consistency
between the experimental
values for $\alpha_s(M_Z)$ and $\sin^2\theta_W(M_Z)$.
Moreover, these models may also avoid the
non--universal contributions
to soft supersymmetry--breaking scalar masses 
in the MSSM that  may arise from
the moduli sector or from charges under an anomalous
$U(1)$ symmetry, leading to well known problems
with flavor--changing neutral currents.
Due to their $Z_2\times Z_2$ orbifold structure,
in the free--fermion models
these contributions may be universal in flavor~\cite{fp},
enhancing the phenomenological viability of these models.
Finally, the symmetries related to the $Z_2\times Z_2$
orbifold structure, which are not present in ordinary GUTs,
may be important for understanding the proton stability \cite{jp,efn}.

The above remarks motivate the need
to understand better the general structure
of the realistic free--fermion models,
particularly in the context of the recent
progress in understanding the nonperturbative 
generalization of weakly--coupled string theories~\cite{reviewsmf}.
A key problem is to study how the advances in understanding
nonperturbative aspects of string theories
may be applied to the free--fermion models.
The purpose of this paper is to take a first step towards
this goal.

Most of the discussion on nonperturbative
generalizations of string compactifications
are in the context of geometrical compactifications, {\it i.e.},
Calabi--Yau manifolds including $K3$ and toroidal compactifications.
The free-fermion models
are however constructed at a fixed point in the moduli space,
and the direct notion of a compactified manifold is lost.
Therefore, a desirable step in advancing
our understanding of free-fermion models to
the nonperturbative regime is to find the
Calabi--Yau three fold which correspond to the
compactification underlying the free--fermion models.
A well known connection exist between Calabi--Yau
compactifications and their realization in terms of
Landau--Ginzburg potentials. In this paper
we therefore seek to find a Landau--Ginzburg
model that may correspond to the $Z_2\times Z_2$
orbifold model at the free--fermion point in the Narain
moduli space. We will also find a natural candidate in terms of a
Calabi--Yau orbifold. In the toroidal language, the $Z_2\times Z_2$
orbifold model at the free--fermion point differs from the
one traditionally discussed in the literature. The difference
is that at the free--fermion point the symmetry of the
compactified lattice is enhanced to $SO(12)$, whereas for
the traditional $Z_2\times Z_2$ orbifold one compactifies
on a $(T_2)^3$ lattice with $SO(4)^3$ symmetry.
This distinction results in the two models having different
Euler characteristics. Whereas the $Z_2\times Z_2$ orbifold
on the $SO(4)^3$ lattice produces $(h_{11},h_{21})=(51,3)$,
that on the $SO(12)$ lattice produces 
$(h_{11},h_{21})=(27,3)$~\cite{foc,efn}.
Therefore, our aim here is to find a Landau--Ginzburg
realization of the $Z_2\times Z_2$ orbifold on $SO(12)$
lattice with $(h_{11},h_{21})=(27,3)$.
In the course of seeking this model
we also examine some of the features of the
Landau--Ginzburg formalism.
As is well known, the \LG~models~\cite{m,vw}
are obtained by specifying a superpotential $W$
and modding by some discrete symmetries of the potential.
To construct a consistent string theory from the \LG~
potential one has to project on states with integral
left-- and right--moving charges and to include in
the spectrum the states from the twisted sector.
One twist that must be included for obtaining
a consistent string theory is that by the scaling
symmetry of the \LG~potential. The novel feature that 
we will examine here is orbifolding of \LG~potentials
by a twist that closes on the scaling symmetry.
This is therefore an extended method for obtaining consistent
\LG~models. In the course of searching for our specific
\LG~orbifold, we will also find new connections between the
different \LG~models reminiscent of the mirror--symmetry phenomenon.

\setcounter{footnote}{0}
\section{The Free--Fermion $Z_2\times Z_2$ Orbifold}

In the free-fermion formulation~\cite{fff}, a model
is defined by a set of boundary condition basis
vectors, together with the related one--loop
GSO projection coefficients, that are constrained
by the string consistency constraints. The basis vectors,
$b_k$, span a finite additive group, $\Xi=\sum_k n_kb_k$
where $n_k=0,\cdots,N_{z_k}-1$.
The physical states in the Hilbert space of
a given sector $\alpha\in\Xi$ are obtained
by acting on the vacuum with bosonic and fermionic
operators and by applying the generalized GSO
projections. The $U(1)$ charges $Q(f)$ corresponding
to the unbroken Cartan generators of the four-dimensional gauge group,
which are in one-to-one correspondence
with the $U(1)$ currents ${f^*}f$ for each complex fermion f,
are given by:
\beq
{Q(f) = {1\over 2}\alpha(f) + F(f)}
\label{qf}
\eeq
where $\alpha(f)$ is the boundary condition of the world--sheet fermion $f$
in the sector $\alpha$, and
$F_\alpha(f)$ is a fermion-number operator that takes the value $+1$ for
each mode of $f$, and the value $-1$ for each mode of $f^*$,
if $f$ is complex. For periodic fermions, which have
$\alpha(f)=1$, the vacuum
must be a spinor in order to represent the Clifford
algebra of the corresponding zero modes. For each periodic complex fermion
$f$, there are two degenerate vacua ${\vert +\rangle},{\vert -\rangle}$,
annihilated by the zero modes $f_0$ and
${{f_0}^*}$, respectively, and with fermion numbers  $F(f)= \pm 1$.

Realistic models in this free-fermion formulation are generated by
a suitable choice of boundary-condition basis vectors for all world--sheet
fermions, which may be constructed in two stages. The first stage consists
of the NAHE set \cite{fsu5,nahe} of five boundary condition basis
vectors, $\{{{\bf 1},S,b_1,b_2,b_3}\}$.
After generalized GSO projections over the NAHE set, the residual gauge
group
is $SO(10)\times SO(6)^3\times E_8$ with $N=1$ space--time
supersymmetry~\footnote{The vector $S$ in this NAHE set is the
supersymmetry generator, and the superpartners of
the states from a given sector $\alpha$ are obtained from the sector
$S+\alpha$.}. The space--time vector bosons that generate the gauge group
arise from the Neveu--Schwarz sector and from the sector $1+b_1+b_2+b_3$.
The Neveu--Schwarz sector produces the generators of
$SO(10)\times SO(6)^3\times SO(16)$. The sector $1+b_1+b_2+b_3$
produces the spinorial 128 of $SO(16)$ and completes the hidden-sector
gauge group to $E_8$.
The vectors $b_1$, $b_2$ and $b_3$ correspond to the three twisted
sectors in the corresponding orbifold formulation, and produce
48 spinorial 16-dimensional representations of $SO(10)$, sixteen each from
the sectors $b_1$, $b_2$ and $b_3$.

The second stage of the basis construction consists of adding three
more basis vectors to the NAHE set, corresponding
to Wilson lines in the orbifold formulation, whose general
forms are constrained by string consistency conditions such as
modular invariance, as well as by space-time supersymmetry.
These three additional vectors are needed to reduce the number of
generations
to three, one from each of the sectors $b_1$, $b_2$ and $b_3$.
The details of the additional basis vectors distinguish between different
models
and determine their phenomenological properties.
This
construction produces a large
number of three-generation models with
different phenomenological characteristics,
some of which are especially appealing.
This class of models corresponds to $Z_2\times Z_2$
orbifold compactification at the maximally-symmetric
point in the Narain moduli space~\cite{Narain}. The emergence
of three generations is correlated with the
underlying $Z_2\times Z_2$ orbifold structure.
Detailed studies of specific models have
revealed that these models may
explain the qualitative structure of the
fermion mass spectrum~\cite{spectrum} and could form the basis of
a realistic superstring model.
We refer the interested reader to several
review articles which summarize the phenomenological
studies of this class of models \cite{ffmreviews}.

The NAHE set is common in a large class of three--generation 
free--fermion models, and its characteristic properties
are reflected in many of the phenomenological
properties of these
models. Therefore, it is important to extend our
understanding of this basic set and its properties.
This set corresponds
to a $Z_2\times Z_2$ orbifold compactification
of the weakly-coupled ten-dimensional heterotic string, and the basis
vectors $b_1$, $b_2$ and $b_3$ correspond to the three twisted
sectors of these orbifold models. To see this correspondence,
we add to the NAHE set the basis vector
\beq
X=(0,\cdots,0\vert{\underbrace{1,\cdots,1}_{{\bar\psi^{1,\cdots,5}},
{\bar\eta^{1,2,3}}}},0,\cdots,0)~.
\label{vectorx}
\eeq
with the following choice of generalized GSO projection coefficients:
\begin{equation}
       c\left( \matrix{X\cr \b_j\cr}\right)~=~
      -c\left( \matrix{X\cr S\cr}\right) ~=~
       c\left( \matrix{X\cr \bone \cr}\right) ~= ~ +1~.
\label{Xphases}
\end{equation}
This set of basis vectors produces models with an
$SO(4)^3\times E_6\times U(1)^2\times E_8$ gauge group
and $N=1$ space--time supersymmetry. The matter fields
include 24 generations in 27 representations of
$E_6$, eight from each of the sectors $b_1\oplus b_1+X$,
$b_2\oplus b_2+X$ and $b_3\oplus b_3+X$.
Three additional 27 and $\overline{27}$ pairs are obtained
from the Neveu--Schwarz $\oplus~X$ sector.

The subset of basis vectors
\beq
\{{\bf1},S,X,I={\bf1}+b_1+b_2+b_3\}
\label{neq4set}
\eeq
generates a toroidally-compactified model with $N=4$ space--time
supersymmetry and $SO(12)\times E_8\times E_8$ gauge group.
The same model is obtained in the geometric (bosonic) language
by constructing the background fields which produce
the $SO(12)$ Narain lattice~\cite{Narain,foc}, taking the metric
of the six-dimensional compactified manifold
to be the Cartan matrix of $SO(12)$:
\beq
g_{ij}=\left(\matrix{~2&-1& ~0& ~0& ~0& ~0\cr%
-1& ~2&-1& ~0& ~0& ~0\cr~0&-1& ~2&-1& ~0& ~0\cr~0& ~0&-1
& ~2&-1&-1\cr ~0& ~0& ~0&-1& ~2& ~0\cr ~0& ~0& ~0&-1& ~0& ~2\cr}\right)
\label{gso12}
\eeq
and the antisymmetric tensor
\beq
b_{ij}=\cases{
g_{ij}&;\ $i>j$,\cr
0&;\ $i=j$,\cr
-g_{ij}&;\ $i<j$.\cr}
\label{bso12}
\eeq
When all the radii of the six-dimensional compactified
manifold are fixed at $R_I=\sqrt2$, it is easily seen that the
left-- and right--moving momenta
\beq
P^I_{R,L}=[m_i-{1\over2}(B_{ij}{\pm}G_{ij})n_j]{e_i^I}^*
\label{lrmomenta}
\eeq
reproduce all the massless root vectors in the lattice of
$SO(12)$,
where in (\ref{lrmomenta}) the $e^i=\{e_i^I\}$ are six linearly-independent
vectors normalized: $(e_i)^2=2$.
The ${e_i^I}^*$ are dual to the $e_i$, and
$e_i^*\cdot e_j=\delta_{ij}$. The momenta $P^I$ of the compactified
scalars in the bosonic formulation can be seen to coincide
with the $U(1)$ charges of the unbroken Cartan generators
of the four dimensional gauge group (\ref{qf}). 

Adding the two basis vectors $b_1$ and $b_2$ to the set
(\ref{neq4set}) corresponds to the $Z_2\times Z_2$
orbifold model with standard embedding.
The fermionic boundary conditions are translated
in the bosonic language to twists on the internal dimensions
and shifts on the gauge degrees of freedom.
Starting from the Narain model with $SO(12)\times E_8\times E_8$
symmetry~\cite{Narain}, and applying the $Z_2\times Z_2$ twisting on the
internal
coordinates, we then obtain the orbifold model with $SO(4)^3\times
E_6\times U(1)^2\times E_8$ gauge symmetry. There are sixteen fixed
points in each twisted sector, yielding the 24 generations from the
three twisted sectors mentioned above. The three additional pairs of $27$
and $\overline{27}$
are obtained from the untwisted sector. This
orbifold model exactly corresponds to the free-fermion model
with the six-dimensional basis set
$\{{\bf1},S,X,I={\bf1}+b_1+b_2+b_3,b_1,b_2\}$.
The Euler characteristic of this model is 48 with $h_{11}=27$ and
$h_{21}=3$.

This $Z_2\times Z_2$ orbifold, corresponding
to the extended NAHE set at the core of the realistic
free--fermion models,
differs from the one which has usually been
examined in the literature~\cite{z2mf}.
In that orbifold model, the Narain
lattice is $SO(4)^3$, yielding a $Z_2\times Z_2$ orbifold model
with Euler characteristic equal to 96, or 48 generations,
and $h_{11}=51$, $h_{21}=3$.

In more realistic free-fermion models, the vector $X$
is replaced by the vector $2\gamma$ with periodic
boundary conditions for $\{{\bar\psi}^{1,\cdots,5},
{\bar\eta}^{1},{\bar\eta}^{2},{\bar\eta}^{3},{\bar\phi}^{1,\cdots,4}\}$
and anti--periodic for all the other world--sheet fermions.
This modification has the consequence of producing a
toroidally-compactified model with $N=4$ space--time supersymmetry and
gauge group $SO(12)\times SO(16)\times SO(16)$.
The $Z_2\times Z_2$ twisting breaks the gauge symmetry to
$SO(4)^3\times SO(10)\times U(1)^3\times SO(16)$.
The orbifold twisting still yields a model with 24 generations,
eight from each twisted sector,
but now the generations are in the chiral 16 representation
of $SO(10)$, rather than in the 27 of $E_6$. The same model can
be realized with the set
$\{{\bf1},S,X,I={\bf1}+b_1+b_2+b_3,b_1,b_2\}$,
projecting out the $16\oplus{\overline{16}}$
from the sector $X$ by taking
\beq
c{X\choose I}\rightarrow -c{X\choose I}.
\label{changec}
\eeq
This choice also projects out the massless vector bosons in the
128 of $SO(16)$ in the hidden-sector $E_8$ gauge group, thereby
breaking the $E_6\times E_8$ symmetry to
$SO(10)\times U(1)\times SO(16)$.
This analysis confirms that the $Z_2\times Z_2$ orbifold on the
$SO(12)$ Narain lattice is indeed at the core of the
realistic free--fermion models.

\setcounter{footnote}{0}
\section{Landau--Ginzburg Construction}

Our aim in this section is to find the Landau--Ginzburg potentials
that produce the same data as the free--fermion $Z_2\times Z_2$
orbifold. We will also find a realization in terms of a Calabi--Yau
orbifold. 
Let us briefly summarize the main steps in the construction
of \LG~ string models \cite{lg}. 
The $N=2$ \LG~ models are defined by a nondegenerate
quasi--homogeneous superpotential $W$ of degree $d$,
\beq
W(\lambda^{n_i}X_i)=\lambda^d W(X_i)
\label{qhsuper}
\eeq
where $X_i$ are chiral superfields, $q_i=n_i/d$ are their
left and right charges under the $U(1)_{J_0}$ current
of the $N=2$ algebra, and the central charge is given by
$c=3\sum_i(1-2q_i)$. For a string vacuum with space--time
supersymmetry we project to integral $U(1)$ charges by taking
the orbifold $W(X_i)/j$, where $j=e^{i2\pi J_0}$,
and need to include the twisted sectors in order to
respect modular invariance. We can take other twistings of
the original Landau--Ginzburg potential by taking
the orbifold $W/G$, where $G$ is a symmetry group
of the non--renormalized $N=2$ superpotential $W(X_i)$,
and can be an arbitrary subgroup of the full
group of symmetries of $W$. The complete Hilbert
space in the orbifold \LG~ theory, 
${\cal H}={\oplus}P_h{\cal H}^h,\, h\in G$ 
then contains
all the states from the untwisted and twisted sectors,
subject to the projection $g\vert\chi\rangle=\vert\chi\rangle$
for every $g\in H$, where $H$ is the center of $G$. 
The states in the $(c,c)$ ring
(the (NS,NS) untwisted sector) are built from the NS vacuum
by monomials in the chiral superfields $X_i$ modulo
setting $\partial_i W=0$, and are thus written as,
\beq
{\prod_i}(X_i)^{(\ell_i)}\vert 0\rangle
\label{nsstates}
\eeq
subject to the condition $\sum_i \ell_iq_i\in Z$
and $\partial W/\partial X_i=0$. 
In the sum over the twisted sectors we have to include
all the sectors up to inequivalent classes of $G$.
We remark that in obtaining the $Z_2\times Z_2$
\LG~ model with $(h_{11},h_{21})=(27,3)$ we will use
a twist that closes on the scaling projection $j$. That is,
for example, $g^2=j^2$. The sum over the twisted sectors then
excludes the sectors containing $g^2$ and $g^3$, which are
equivalent to $j^2$ and $j^2g$.

In the twisted sectors the chiral superfields $X_i$ are twisted
by $X_i(2\pi)=h_i^jX_j(0)$. The chiral superfields are rotated to
a basis in which $h$ is diagonal $h=\delta_i^j{\rm exp}(2\pi i{\Theta_i^h})$.
The charges and degeneracies in the twisted sectors are
then obtained by computing the charge of the lowest--charge
state in the $h$--twisted Ramond sector, $\vert 0\rangle_R^h$, given by
\beq
{J_0\choose{\bar J}_0}\vert 0\rangle_R^h=
\left(\pm \sum_{\Theta_i^h\notin Z}(\Theta_i^h-\left[\Theta_i^h\right]
-1/2)+\sum_{\Theta_i^h\in Z}(q_i-1/2)\right)\vert 0\rangle_R^h
\label{qrs}
\eeq
where $[x]$ denotes the greatest integer smaller than $x$,
and acting on the Ramond ground states with the untwisted
fields 
\beq
{\prod_{\Theta_i^h\in Z}}(X_i)^{(\ell_i)}\vert 0\rangle_R^h
\label{rstates}
\eeq
to produce the twisted integrally charged states.
Using spectral flow on the left and right charges, with $(c/6;c/6)$,
we obtain the charges in the $(c,c)$ ring of the twisted sectors.
At the final step the projections imposed by all $g\in G$ are
applied to produce the modular--invariant spectrum. 

We now turn to the class of \LG~models that reproduces the data
of the desired orbifold models. 
It is well known that the $T^2$ torus has the polynomial representation
\beq
W=X_1^4+X_2^4+X_3^2 
\label{t2}
\eeq
modded by the $Z_4$ scaling symmetry. Note that the complex structure
of the torus is fixed, $\tau=i$ or else the torus would not have a
$Z_4$ symmetry~\footnote{The \LG~potential for a generic torus is
given by $W=X_1^6 + X_2^3 + X_3^2$.}. Here the field $X_3$ is a trivial
superfield as its conformal charge is zero, and is used to mimic the
geometrical representation of $T^2$ for $W=0$.
The starting point for our construction is therefore the
holomorphic superpotential given by
\beq
W=X_1^4+X_2^4+X_3^2+X_4^4+X_5^4+X_6^2+X_7^4+X_8^4+X_9^2
\label{quarticpotential}
\eeq
which is a just a triple sum of the superpotential in (\ref{t2})
and $X_{3,6,9}$ are again  trivial superfields.
The superpotential in (\ref{quarticpotential})
correspond to a super--conformal field theory with
$c=9$. 

In order to construct a consistent string vacuum we need to consider
the \LG-orbifold $W/j$ where $j$ is the scaling symmetry
\beq
Z_4^j~:~(X_1, \ldots, X_9)~\rightarrow~ 
   (i X_1, i X_2, - X_3, i X_4, i X_5, -X_6, 
   i X_7, i X_8, -X_9)~.
\label{jaction}
\eeq
Let us denote this \LG\ orbifold ${\cal M}$.
The next stage in the construction is to mod out the superpotential
by $Z_2$ and $Z_4$ symmetries. The spectrum is then obtained
by analyzing the untwisted and twisted sectors, as outlined above.
We note that the superpotential (\ref{quarticpotential}) (modulo the
trivial superfield  $X_{3,6,9}$)  
has a large symmetry group $Z_4^6$, where
each superfield can be rotated separately by a $Z_4$ twist.

Rather than considering a \LG~orbifold corresponding to the $Z_2\times
Z_2$ orbifold model with 
$(h_{11},h_{21})=(51,3)$ we will first construct the mirror model with
$(h_{11},h_{21})=(3,51)$~\footnote{Since the \LG~orbifold is just a
quotient of a product of minimal models, mirror symmetry can be shown
at the level of conformal field theory~\cite{gp}.}. To obtain the mirror 
we take ${\cal M}/(Z_2^A\times Z_2^B)$ where
\beqn
&& Z_2^A~:~(X_1,\cdots,X_9)~\rightarrow~ 
   (X_1, X_2,X_3, -X_4, -X_5,X_6,-X_7,-X_8,X_9)~;~~~~~~~~~~\nonumber\\
&& Z_2^B~:~(X_1,\cdots,X_9)~\rightarrow~ 
   (-X_1, -X_2, X_3, -X_4, -X_5,X_6,X_7,X_8,X_9)~.~~~~~~~~~~\label{z2z2513}
\eeqn
This orbifold contains 30 $(1,1)$ states from the untwisted sector,
one from each of the twisted sectors $\Theta^g$,
and six from each of the twisted sectors $\Theta^{j^2g}$,
where $g=(Z_2^A,Z_2^B, Z_2^A\otimes Z_2^B)$, making a total of 51 $(1,1)$
states. It contains three (-1,1) states from each of the twisted sectors
$\Theta^{jg}$, thus reproducing the data of the $Z_2\times Z_2$ orbifold
on the $SO(4)^3$ lattice. We note that the cyclic permutation symmetry
between the twisted and untwisted spectrum, the characteristic
property of the $Z_2\times Z_2$ orbifold, is exhibited in this model.

The mirror of the
$Z_2\times Z_2$ \LG~ orbifold model, with $(h_{11},h_{21})=(3,27)$,
can be reproduced in several ways. 
One option is to take the
twist
\beqn
&& Z_4^A~:~(X_1,\cdots,X_9)~\rightarrow~ 
   (X_1,X_2,X_3,i X_4,-i X_5,X_6,i X_7,-i X_8,X_9)~;~~~~~~
					\nonumber\\
&& Z_4^B~:~(X_1,\cdots,X_9)~\rightarrow~ 
   (i X_1,-i X_2,X_3,i X_4,-i X_5,X_6,X_7,X_8,X_9)~.~~~~~~
					\label{z4z4273}
\eeqn
and the three trivial superfields
are kept fixed, {\it i.e.} $(X_3,X_6,X_9)\rightarrow (X_3,X_6,X_9)$.
This choice however mixes the action
of the $Z_4$ and $Z_2\times Z_2$ twists. Therefore, the cyclic
permutation symmetry of the $Z_2\times Z_2$ orbifold is not transparent
with this choice. A more elegant choice is obtained by
taking the $Z_2^A\times Z_2^B$ twist of (\ref{z2z2513})
and adding the following $Z_4$ twist,
\beqn
&& Z_4^{w}~:~(X_1,\cdots,X_9)~\rightarrow~ 
   (i X_1,-i X_2,X_3,i X_4,-i X_5,X_6,i X_7,-i X_8,X_9)~.~~~~~~~
					\label{z2z2z4273}
\eeqn
Here again the trivial superfields are kept fixed.
This twist closes on the scaling symmetry, $w^2=j^2$.
Therefore, in summing over the twisted spectrum we only
have to sum up to inequivalent classes of the twist group.
That is, twisted sectors that contain $w^2$ or $w^3$
are excluded as they are equivalent to $j^2$ and $jw^2$.

An alternative way of obtaining the same model is to
note that the $Z_4$ twist in (\ref{z2z2z4273}) acts
only as a $Z_2$ modding, as it closes on the $Z_4$ scaling projection.
Thus, the $Z_2$ quotient is generated by taking the 
product $Z_4^j\otimes Z_4^w$.
Then, taking the $Z_2^A\times Z_2^B$ twist of eq. (\ref{z2z2513})
and adding the above $Z_2$ twist:
\beqn
&& Z_2^{w}~:~(X_1,\ldots, X_9)~\rightarrow~ 
   (-X_1, X_2, -X_3, -X_4, X_5, -X_6, -X_7, X_8, -X_9)~,~~~~~~~
					\label{z2z2273}
\eeqn
we have the \LG\ orbifold ${\cal M}/(Z_2^A\times Z_2^B\times Z_2^\omega)$.
Here the three trivial superfields are twisted by the
$Z_2^w$ twist.
We then obtain in this \LG~ orbifold model 18 (1,1) states
from the untwisted sector, one (1,1) state from each of the sectors
$\Theta^g$ with $g=(Z_2^A,Z_2^B, Z_2^A\otimes Z_2^B)$,
and two (1,1) states from each of the twisted sectors $\Theta^{j^2g}$,
producing
a total of 27 (1,1) states, as desired. Three (-1,1) states
are obtained, one from each of the twisted sectors
$\Theta^{jg}$, thus reproducing the data of the $Z_2\times Z_2$ orbifold
on the $SO(12)$ lattice. We note that the cyclic permutation
symmetry between the twisted sectors is retained in this model.
In fact, the model is seen to be obtainable from the $(51,3)$
$Z_2\times Z_2$ \LG~ orbifold model by the action of the
$Z_4^w$ twist (\ref{z2z2z4273}), or by the action of
the $Z_2^w$ twist (\ref{z2z2273}). The action of this twist
is to reduce the number of (1,1) untwisted states from 30 to 18
and the number of twisted states from the sectors $\Theta^{j^2g}$
from 18 to 6, thus reducing the total number of (1,1) states from
51 to 27, as needed. The projection (\ref{z2z2273}) is
therefore seen to connect between the mirrors of the $(51,3)$
and $(27,3)$ $Z_2\times Z_2$ \LG~ orbifold models, respectively.

So far we have only constructed the mirror models of the relevant
$Z_2\times Z_2$ orbifolds. The \LG~orbifolds for the $Z_2\times Z_2$
orbifolds can, however, also be constructed. In particular,  
we note that the operation of the twist in (\ref{z2z2z4273}) or
(\ref{z2z2273}) also works
for these models. The (51,3) model
is obtained by the following twists,
${\cal M}/(Z_4^A\times Z_4^B)$:
\beqn
&& Z_4^A~:~(X_1, \ldots, X_9 )~\rightarrow~ 
 (X_1,X_2, X_3,i X_4,i X_5, X_6, i X_7,
 i X_8, X_9)~;~~~~~~~~~ 
					\nonumber\\
&& Z_4^B~:~(X_1, \ldots, X_9 )~\rightarrow~ 
 (i X_1,i X_2, X_3,i X_4,i X_5, X_6, X_7, X_8, X_9 )~.~~~~~~~~~
					\label{z4z4351}
\eeqn
This \LG~ model has 3 (1,1) states from the untwisted sector
and 51 (-1,1) states from the twisted sectors. This model
therefore contains the spectrum  of the (51,3) model.
Just as for its mirror,
the states from the (un)twisted sectors can be shown to satisfy the cyclic
permutation symmetry characteristic of the $Z_2\times Z_2$ orbifold.
Adding to this model the twist in (\ref{z2z2z4273}) or (\ref{z2z2273})
then produces
the model with 3 (1,1) states from the untwisted sector
and 27 (-1,1) states from the twisted sectors. The projection
(\ref{z2z2z4273}) (or (\ref{z2z2273}))
is therefore observed to
commute with the mirror symmetry operation. In the free--fermion
$Z_2\times Z_2$ orbifold models the mirror symmetry transformation
operates by taking the GSO coefficients $c(b_i,b_j)\rightarrow-c(b_i,b_j)$
and can be seen in the orbifold models to operate by the choice
of the discrete torsion. The projection (\ref{z2z2273})
may therefore be regarded as a new geometrical operation
that connects between models with different Euler
characteristic.
It is of great interest to examine
how this can be interpreted in a more geometrical language. We now
turn to this point.

The projection (\ref{z2z2273}), that we used to obtain the $(27,3)$
$Z_2\times Z_2$ \LG~ orbifold model from the $(51,3)$ one, can be
given a very natural geometric interpretation. Consider $(T^2)^3$
given in terms of an intersection of three quartic equations,
$p_i(x)=0$, 
\beq
p_i = (x_{1+3i})^4 + (x_{2+3i})^4 + (x_{3+3i})^2=0,\quad i=0,1,2~.
\label{tori}
\eeq
The $Z_2\times Z_2$ orbifold with $(h_{11},h_{21})=(51,3)$ is then
represented by  
\beqn
&&Z_2^1~:~(x_1,\ldots,x_9)\to (x_1,x_2,x_3,-x_4,x_5,x_6,-x_7,x_8,x_9)~~~~~~~~~ 
					\nonumber\\
&&Z_2^2~:~(x_1,\ldots,x_9)\to (-x_1,x_2,x_3,-x_4,x_5,x_6,x_7,x_8,x_9)~.~~~~~~~ 
\label{ztwoztwo}					
\eeqn
There are 16 fixed tori in each of the twisted sectors due to
$(Z_2^1,Z_2^2,Z_1\otimes Z_2^2)$, each contributing 16 $(1,1)$-forms.
We then obtain the $(27,3)$ $Z_2\times Z_2$ orbifold by the $Z_2^w$ quotient
(\ref{z2z2273}). 
We note that $Z_2^w$ is freely acting, {\it i.e.}, there is no 
fixed-point set. Moreover, because of this fact, the new Calabi--Yau
manifold
has a non-trivial homotopy class, $\Pi_1=Z_2$. 

The projection (\ref{z2z2z4273})
introduced the novel feature that it closes on the scaling
symmetry $j$ of the \LG~ superpotential (\ref{quarticpotential}).
We now elaborate further on this type of projection
by studying the consequences of similar projections in 
two and four compactified dimensions. In the two--dimensional
case, the \LG~ superpotential is that of (\ref{t2}).
In this case the requested projection is simply
\beq
(X_1,X_2)~\rightarrow~(i X_1, -i X_2)
\label{z4t2}
\eeq
The original \LG~ theory $W/\{j\}$ correspond to a $T_2$ torus
and therefore one (1,1) state from the untwisted sector,
$X_1^2X_2^2$. In this case we note that the projection $g$
in (\ref{z4t2}) leaves this state invariant, and does not give
rise to any additional states. Therefore in this case the
\LG~ theories $W/\{j\}$ and $W/\{j,w\}$ are identical. This
in fact may have been expected as in one complex dimension
the $T_2$ torus is the only Ricci--flat manifold. 

We now turn to the four--dimensional case. The special projection is
\beq
(X_1,X_2,X_3,X_4)~\rightarrow~(i X_1, 
			-i X_2, i X_3,-i X_4)
\label{z4t4}
\eeq
In this case the \LG~ theory $W/\{j\}$ admits 19 (1,1) states
from the untwisted sector and one (1,1) state from the sector
$j^2$. In the \LG~ theory  $W/\{j,w\}$ the effect of the $w$ projection
is to project eight of the (1,1) states from the untwisted
sector. The three twisted sectors $\{j^2,w,j^2w\}$ each give
one (1,1) state, whereas the two twisted sectors $\{jw,j^3w\}$
each give 3 (1,1) states. Therefore, in this case we note
that the two theories $W/\{j\}$ and $W/\{j,w\}$ again reproduce
the same spectrum. 

Finally, we turn to six dimensions, where the \LG~
potential is given by (\ref{quarticpotential}).
In this case the theory $W/\{j\}$ gives rise to 90 
(1,1) states from the untwisted sector. The theory
$W/\{j,w\}$ with $w$ given by (\ref{z2z2273})
produces 48 (1,1) states from the untwisted sector
and three (1,1) states from each of the twisted
sectors $\{jw,j^3w\}$, giving a total of 54 (1,1) states.
Therefore we see that while in the one and two complex--dimensional cases
the effect of the special $w$ projection is
trivial, in the case of three complex dimensions
new features are encountered which modify the original
\LG~ theory $W/\{j\}$. It is therefore of interest to
examine further the geometrical representation of
this projection in Calabi--Yau manifolds. 

We have found above two related ways to connect the (51,3) $Z_2\times Z_2$
orbifold to the (27,3) one. The first utilized the $Z_4^w$ projection
 (\ref{z2z2z4273}), whereas the second utilized the $Z_2^w$
projection (\ref{z2z2273}). The two projections
are related by $Z_2^w\equiv Z_4^j\otimes Z_4^w$.
Therefore, the substitution of (\ref{z2z2273}) for
(\ref{z2z2z4273}) corresponds to choosing different
elements to generate the same group.
However, the difference
between the two projections, which is interesting from
geometrical view of these potentials, is that in the first
case, eq. (\ref{z2z2z4273}), the trivial superfields are left invariant
by the action of $Z_4^w$,
whereas with the second choice (\ref{z2z2273}), the trivial superfields
are twisted by the action of $Z_2^w$. {}From
a geometric point of view, the addition of the trivial
superfields is very natural, as it mimics the structure of
$(T^2)^3$. 
In terms of \LG\ orbifold there is no need to have the extra trivial
superfields, making the $Z_4^w$ quotient the more natural operation.
We note, however, that the fact that twisting
of trivial superfields results in new models makes the prospects
of classifying \LG~ orbifolds with $c=9$ more problematic (see
however~\cite{ks}),  as
we can add any number of such fields to the superpotential.
However, the equivalence of such a twist to a twist that closes on the
scaling symmetry might facilitate the classification.

\setcounter{footnote}{0}
\section{Discussion and Conclusions}

The realistic free--fermion models have offered to date the 
most promising prospect for producing a realistic superstring
model. The main reasons to believe that the true string
vacuum is located in the vicinity of this class of
string vacua is not due to the detailed properties of
one of the specific solutions which have been constructed
to date, but rather due to the underlying gross structure
of these models and their general properties. These
include the correlation of three generations with the
underlying compactification and the natural standard $SO(10)$
embedding of the weak hypercharge. These models
naturally give rise to flavor--universal untwisted moduli,
which is not expected to be the case in a generic superstring model.
Finally, these models naturally give rise to symmetries that
may be required in order to maintain the proton stability.
Thus, we see that the realistic free--fermion models naturally
contain the necessary ingredients that a realistic string model
must contain. The realistic nature of the free--fermion models
arises for a very basic reason. They are based on an
underlying $Z_2\times Z_2$ orbifold compactification.
It is therefore of vital importance to extend our
understanding of this particular class of orbifold
compactifications.

The nonperturbative extension of the weakly coupled free--fermion
models will surely offer important further insight to their properties.
For this purpose we may attempt to use the various $F$-- (and $M$--)
theory dualities that have been recently discovered.
The key point
in $F$~theory, compactified to six dimensions, is that
the models admit an elliptic fibration in which a Calabi--Yau
threefold is identified as a two complex--dimensional base
manifold $B$ with an elliptic fiber. 
A class of elliptically fibered Calabi--Yau three--folds
which can be written as $K_3\times T_2$ modded out with an appropriate
$Z_2$ symmetry, has been
analyzed by Voisin~\cite{voisin} and Borcea~\cite{borcea} (see also 
\cite{nikulin}). Among these
models are those which have an orbifold interpretation. In particular
there are $Z_2\times Z_2$ orbifolds
with different $h_{11}$ and $h_{21}$. For example, in~\cite{z2mf}
the models with $(h_{11},h_{21})=\{(11,11);(19,19);(51,3)\}$
have been examined in connection with $F$--theory compactifications.
These models are constructed as orbifolds of $(T^2)^3$ with 
a generic choice of
complex structure for the tori. 
However, the above $Z_2\times Z_2$ orbifold models, and many more, can also be
obtained from the quartic \LG~ superpotential, namely
(\ref{quarticpotential}) with a possible suitable
choice of an additional $Z_4$ projection. Because of the choice of a
square lattice for the tori,
rather than having the full $SL(2,Z)$ action on the elliptic fiber, we
are restricted to a one-parameter family which leaves $\tau=i$
invariant. This could, however, still be a good $F$-theory vacuum.
Alternatively, these models can be studied in five dimensions in terms
of compactifications of $M$~theory.
In this regard, the key property
of the $Z_2\times Z_2$ orbifold that should serve as the basic
guide in seeking its possible nonperturbative extension is its basic
characteristic property, namely the cyclic permutation symmetry
of the twisted and untwisted sectors.
It may not be a coincidence that this characteristic property of the
$Z_2\times Z_2$
orbifold, which is clearly of vital phenomenological importance,
may also serve as the guiding light in the search
for the nonperturbative extension of the phenomenologically
realistic free--fermion models.

\bigskip
\medskip
\leftline{\large\bf Acknowledgments}
\medskip

We are pleased to thank David Morrison and Cumrun Vafa for discussions.
This work was supported in part by the Department of Energy
under Grants No.\ DE-FG-05-86-ER-40272 and DE-FG03-95-ER-40917.
The work of P.B. was supported in part by the National
Science Foundation grant NSF  PHY94-07194.

\vfill\eject

\bibliographystyle{unsrt}

\end{document}